\documentclass{article}
\usepackage[utf8]{inputenc}
\usepackage[a4paper,top=2.5cm,bottom=3cm,left=2.5cm,right=2.5cm,marginparwidth=2cm]{geometry}
\usepackage[hidelinks]{hyperref}
\usepackage{amsmath, amssymb, amsthm, bbm, graphicx, caption, tikz, subcaption, siunitx, enumerate, epstopdf}
\newcommand{\tr}{\mathrm{tr}}
\newcommand{\ii}{\mathrm{i}}

\theoremstyle{definition}

\numberwithin{equation}{section}
\numberwithin{thm}{section}

\title{Hofstadter butterflies in phononic structures: commensurate spectra, wave localization and metal-insulator transitions}
\author{Bryn Davies$^{(1)}$ \hspace{1cm} Lorenzo Morini$^{(2)}$\thanks{lorenzo.morini@unica.it}\\
$^{(1)}${\normalsize {Mathematics Institute, University of Warwick, Coventry CV4 7AL, UK.}}\\
$^{(2)}${\normalsize {Department of Mechanical, Chemical and Materials Engineering, University of Cagliari,}}\\
{\normalsize {via Marengo 2, 09123 Cagliari, Italy.}} 
  }
 \date{}
\usepackage{todonotes}

\begin{document}
\maketitle

\begin{abstract}
 We present a new simple and easy-to-implement one-dimensional phononic system whose spectrum exactly corresponds to the Hofstadter butterfly when a parameter is modulated. The system consists of masses that are coupled by linear springs and are mounted on flexural beams whose cross section (and, hence, stiffness) is modulated. We show that this system is the simplest version possible to achieve the Hofstadter butterfly exactly; in particular, the local resonances due to the beams are an essential component for this. We examine the various approaches to producing spectral butterflies, including Bloch spectra for rational parameter choices, resonances of finite-sized systems and transmission coefficients of sections of finite length. For finite-size systems, we study the localisation of the modes by calculating the inverse participation ratio, and detect a phase transition characterised by a critical value of the stiffness modulation amplitude, where the state of the system changes from mainly extended to localised, corresponding to a metal-insulator phase transition. The obtained results offer a practical strategy to realize experimentally a system with similar dynamical properties. The transmission coefficient for sections of finite length is benchmarked through the comparison with Bloch spectra of the same finite-sized systems. The numerical results for the transmission spectra confirms the evidence of a phase transition in the dynamical state of the system. Our approach opens significant new perspectives in order to design mechanical systems able to support phase transitions in their vibrational properties. \\
%Although the transmission coefficient gives the lowest resolution image, this offers a practical %strategy to reproducing these results experimentally.\\

% Transmission coefficient is the easy one to realise in practice (for this simple mechanical structure).

% The contributions of this paper are (1) presenting a simple and easy-to-implement mechanical/phononic system whose spectrum exactly corresponds to the Hofstadter butterfly, (2) showing that this is the simplest possible system that can achieve this (in particular, you need the local resonances), (3) surveying the different ways that spectral butterflies can be computed, based on this system (Bloch curves, finite eigenvalues and transmission spectra), (5) giving a practical strategy to realise the metal-insulator transition in mechanical systems
\noindent
Keywords: Hofstadter butterfly spectrum, Quasiperiodic modulation, Localised mode, Inverse participation ratio, Metal-insulator transition.

\end{abstract}

% \begin{enumerate}
%     \item Show that the proposed metamaterial has a spectrum that is given exactly by the Hofstadter butterfly. (Make the picture three different ways: (1) Floquet-Bloch for just the rational number; (2) Hamiltonian eigenvalues for a finite system with edges; (3) transmission coefficient for a finite-sized array.)
%     \item Plot how the periodic approximants behave. (Prove that there are super band gaps, at least in the case that $\xi$ is rational. They also seem to exist for irrational $\xi$... can we prove or explain it? As the limit of rational $\xi$? Perhaps random methods would work well here?)
%     \item Can we study irrational values of $\xi$ somehow? Perhaps as the limit of rational approximants?
% \end{enumerate}

% \todo[inline]{To do: Make the Floquet Bloch pictures clearer, maybe 3D visualisation (MATLAB 3D line plot). Make the positions of the finite modes clearer (arrows with the circles). Make the plots with $\omega$ on the y axis instead. Compute the transmission coefficient (try doing just $1/M_{22}$ as a first guess). Move the Appendix A argument to the main part of the paper? Maybe compute the transmission coefficient and some modes to show the metal-insulator transition?}

\section{Introduction}

The recent growth in the exploration of quasiperiodic wave systems has been driven by their ability to support complex, exotic and often fractal spectra. These have the potential to expand the design space for dynamical metamaterials and meta-structures beyond classical periodic configurations, improving their potential in terms of wave control and providing the possibility to exploit different phenomena in future applications, especially for what concerns the so-called topological effects \cite{pal2019topological, xia2020topological}. One of the most famous examples of a fractal spectrum is the Hofstadter butterfly \cite{hofstadter1976energy}, which has been the inspiration for a wealth of recent studies considering the spectra of structured media with incommensurate periodic modulation. These studies span many different physical domains, including quantum mechanics \cite{dean2013hofstadter}, photonics \cite{ye2022hofstadter, manela2010hofstadter}, phononics \cite{gupta2020dynamics, marti2021edge, silva2019phononic, rosa2021exploring, moscatelli2024wave, comi2024some} and vibro-acoustics \cite{richoux2002acoustic, ni2019observation}, but all seek to obtain images of spectra that resemble the Hofstadter butterfly when the modulation parameter is varied. This modulation parameter is usually a geometric parameter which characterises the relative phase of the modulation and captures the system's periodicity or aperiodicity (typically corresponding to rational or irrational values, respectively). The modulated spectra typically display fractal and self-similar properties, qualitatively reminiscent of Hofstadter's famous butterfly image \cite{hofstadter1976energy}.

In his seminal 1976 paper on the energy levels of Bloch electrons in irrational magnetic fields \cite{hofstadter1976energy}, Hofstadter obtained the famous ``butterfly'' spectral pictures by finding the eigenvalues $\mathcal{E}$ of Harper's equation \cite{harper1955single}:
\begin{equation} \label{eq:harper}
    g(n+1)+g(n-1)+\lambda\cos(2\pi\phi n)g(n)=\mathcal{E}g(n), \quad n\in\mathbb{Z},
\end{equation}
for different values of the parameter $\phi$, with $\lambda=2$. This is often stated with a phase-shift parameter within the cosine, but this just has the effect of translating the spectrum, so we omit it here. As we will see further in this work, varying the value of $\lambda$ can alter the nature of the eigensolutions yielding to states of the system which have different dynamical properties with respect to the most widely studied variant of \eqref{eq:harper} corresponding to $\lambda=2$\cite{jitomirskaya1999metal}.

The spectrum of the eigenvalue problem \eqref{eq:harper} has been studied extensively and has led to some longstanding and notoriously challenging problems in spectral theory \cite{aubry1980analyticity, avila2009ten, jitomirskaya1999metal}. However, several simple properties of the spectrum have been well known since Hofstadter's original work. For example, any eigenvalue $\mathcal{E}$ of \eqref{eq:harper} must satisfy $-2-\lambda\leq \mathcal{E}\leq 2+\lambda$ and $\mathcal{E}$ is an eigenvalue if and only if $-\mathcal{E}$ is an eigenvalue. 

The main recent interest from the mechanical metamaterials and phononics community in designing systems with spectra corresponding to the Hofstadter butterfly has been motivated by the possibility to exploit the fractal nature of these spectra to control different wave phenomena and optimize the associated dynamical properties. The key to producing this type of phononic structures is to modulate their parameters sinusodially,  as in \eqref{eq:harper}. Several different quasiperiodic systems have been implemented with Harper-type modulation on the coupling strengths \cite{pal2019topological, gupta2020dynamics, xia2020topological, marti2021edge, silva2019phononic, moscatelli2024wave}. In these examples, the coupling strength is typically modulated by varying the positions of resonators on a homogeneous substrate (such as an elastic plate or beam). This yields pictures that qualitatively resemble the Hofstadter butterfly, in the sense that they have a pattern of many different spectral gaps that are shifted when the modulation parameter $\phi$ is varied. However, for a system to correspond exactly to \eqref{eq:harper} (such that we are able to leverage the whole range of mathematical results that have been developed for this operator), we need to perform ``on-site'' modulations by varying the locally resonant elements themselves. Only a few mechanical systems with on-site modulation have been previously proposed, for example by \cite{ni2019observation, richoux2002acoustic} in arrays of coupled acoustic resonators, by \cite{rosa2021exploring} in a LEGO beam, and by \cite{martinez2018quasiperiodic} in granular chains with local resonances provided by either an external restitution mechanism or an internal single-degree-of-freedom resonance. 

The other important practical question is, given a model for a physical system (with suitable modulation), how to generate images reminiscent of the Hofstadter butterfly spectrum from \eqref{eq:harper}. A common strategy in the literature is to compute the eigenmodes of a finite-sized sample. However, the introduction of boundaries adds additional eigenvalues to the spectrum, corresponding to edge modes \cite{pal2019topological, xia2020topological, silva2019phononic, ni2019observation}. Although these modes, which are localised at one of the boundaries of the system, are interesting in their own right (\emph{e.g.} they have interesting topological properties \cite{pal2019topological, xia2020topological, silva2019phononic, ni2019observation}), it is often desirable to remove them to recover the classical, unpolluted butterfly. Algorithms have been proposed for identifying edge modes (by studying their profiles) and removing them from the spectrum \cite{lian2021open}, but this is computationally expensive for high-resolution plots. In this paper, we will show further how the restriction to rational modulation parameters $\phi$ is sufficient to produce high-resolution butterfly spectra with and is significantly more computationally efficient. This is because any rational $\phi$ will yield a periodic system (albeit potentially with a very large unit cell), meaning conventional methods (\emph{e.g.} transfer matrices) can be used to compute the Bloch spectra. We will also show how both these spectra (finite eigenvalues and Bloch spectra) are reproduced by studying the transmission coefficients of finite-sized sections of the phononic system, with a resolution that increases with the length.

In this work, we present a new design for a simple phononic system that supports Hofstadter butterfly spectra. The system we propose can be implemented on a table-top setup using masses coupled with springs mounted atop beams with varying thickness. We show this systems is minimal, in the sense that it is not possible to remove any elements and still realise Hofstadter's butterfly (in particular, the modulation of the local beam resonances is essential). We will present a comprehensive theoretical exposition of the problem and survey the different strategies for making butterfly pictures (Bloch spectra, finite spectra and transmission spectra), highlighting the differences between the approaches (such as differences in resolution and tendency to produce polluted spectra). For finite-sized systems, we highlight a transition from a banded to an unbanded spectrum where most of the states are grouped in very narrow sets such that it is difficult to define them as a band. By calculating the inverse participation ratio \cite{biddle2012IPR} of the different eigenmodes, we determine a critical value of the amplitude of the beam stiffness modulation associated with a change in the dynamic state of the system, which becomes from mainly extended to localised, similarly to a metal-insulator phase transition. This result, confirmed by the analysis of the transmission spectra, show that the features of relatively simple one-dimensional systems may be profitably exploited for the design of phononic structures supporting phase transitions in their dynamical properties. This could open new perspectives towards the realisation of tunable waveguides operating alternatively as insulators or as vibration conductors.
%Numerical results show that for finite-sized systems the transition from a banded to an %unbanded spectrum where most of the states are grouped in very narrow sets such that it %is difficult to define them as a band is governed by the amplitude of the beams stiffness %modulation. We study the localisation of the modes by calculating the inverse %participation ratio, and detect a phase transition characterized by a critical value of %the stiffness modulation amplitude, where the state of the system from mainly extended to %localised, similarly to a metal-insulator phase transition [2, 3].  

\section{Mechanical system}

\begin{figure}
    \centering
    \includegraphics[width=0.8\linewidth]{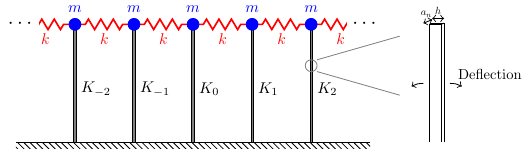}
    \caption{We consider a system of masses that are mounted on Euler-Bernoulli beams and coupled by linear springs. The specific case we consider in this work has identical springs with stiffness $k$ and identical masses $m$ but modulated beams with stiffness $K_n$, which we choose such that the system is described by Harper's equation \eqref{eq:harper}. }
    \label{fig:systemsketch}
\end{figure}

We consider a mechanical system whose spectrum corresponds exactly to the Hofstadter butterfly. The system is composed of a system of masses that are mounted on Euler-Bernoulli beams and coupled by linear springs. We suppose that the masses all have identical mass $m$, the springs all have identical stiffness $k$ and the physical and geometrical properties of the beams are modulated such that the $n$th beam has stiffness $K_n$. This is depicted in Figure~\ref{fig:systemsketch}. 

%\textcolor{red}{
%The modulation of the beam can be achieved by varying the thickness or ...
%}

%\todo[inline]{Add details of how to modulate the beam stiffness. Suggest varying the cross %section, such that $K=EI/l^3$ is varied. Do either square or circular cross section (add a sketch %to figure 1).}

In order to derive the governing equation for this system, we consider the forces acting on a given mass. We neglect any influence of gravity on the system, so the only forces acing on a given mass are due to the springs and the beam. Suppose that the $n$th mass has displacement $u_n(t)$. Then, the spring on the right of the $n$th mass exerts a force $k(u_{n+1}-u_n)$ on the $n$th mass. Similarly, the spring to the left of the $n$th mass exerts a force given by $-k(u_{n}-u_{n-1})$. Finally, the mass also experiences a restorative force due to the deflection of the beam. This force is proportional to its displacement and is given by $- K_n u_n$. Summing these three forces to give the total force on the $n$th mass, we find that Newton's second law gives
\begin{equation}
    m \frac{\mathrm{d}^2}{\mathrm{d}t^2}u_n=-k(u_{n}-u_{n-1})+k(u_{n+1}-u_n)- K_n u_n.
\end{equation}
Let us study the propagation of time-harmonic waves through the system, so that we can write $u_n(t)=g(n)\exp(\ii\omega t)$ for some frequency $\omega\in\mathbb{R}$. Substituting this ansatz gives the discrete difference equation
\begin{equation} \label{eq:difference}
    -\omega^2 m g(n)=-k\big(g(n)-g(n-1)\big)+k\big(g(n+1)-g(n)\big)- K_n g(n).
\end{equation}

We now choose to modulate the stiffness $K_n$ in such a way that \eqref{eq:difference} becomes Harper's equation \eqref{eq:harper}. For this purpose, we introduce the expression $K_n=3EI_n/l^3$, where the Young's modulus $E$ and the length $l$ are assumed to be the same for all beams, whereas the second moment of area $I_n$ is modulated. The modulation of $I_n$ can be performed, for example, for rectangular or elliptical cross sections. For rectangular cross sections, the second moment of area is given by $I_n=a_n h^3/12$, where $a_n$ is the width perpendicular to the direction of deflection and $h$ is the width parallel to the deflection.
% basis parallel to the horizontal axis corresponding to the $n$-th beam and $h$ is the height parallel to the vertical axis.
To achieve the  modulation as in Harper's equation \eqref{eq:harper}, we choose to vary $a_n$ so that
\begin{equation} \label{eq:rectangular}
I_n=\frac{a_n h^3}{12}=\frac{ah^3 }{12}\left(1+b-\cos(2\pi \phi n)\right),
\end{equation}
where $\phi$ and $b$ are non-dimensional parameters satisfying $0<\phi<1$ and $b>0$. Conversely, in the case of elliptical cross sections, $I_n$ is given by
\begin{equation} \label{eq:elliptical}
I_n=\frac{\pi}{4}\alpha_n \beta^3=\frac{\pi \alpha\beta}{4}\left(1+b-\cos(2\pi \phi n)\right),
\end{equation}
where $\alpha_n$ is the semi-axis perpendicular to the direction of deflection and $\beta$ is the semi-axis parallel to the deflection. Substituting the expressions \eqref{eq:rectangular} and \eqref{eq:elliptical} into the formula $K_n=3EI_n/l^3$, we obtain 
\begin{equation} \label{eq:modulation}
K_n= C\left(1+b-\cos(2\pi \phi n)\right),
\end{equation}
where for rectangular and elliptical cross sections the mean stiffness parameter $C$ is given by 
\begin{equation}\label{eq:meanstiffness}
C=\frac{Eah^3}{4l^3} \quad \mathrm{and} \quad C=\frac{3\pi E\alpha\beta^3}{4l^3},
\end{equation}
respectively. It is important to note that the assumption of $b>0$ is made so that $K_n$ is always positive (we have that $b\leq K_n\leq 2+b$ for any $n$ and $\phi$). In this case, \eqref{eq:difference} can be rewritten as 
\begin{equation} \label{eq:diff2}
    g(n+1)+g(n-1)+\frac{C}{k}\cos(2\pi\phi n)g(n) = \left( 2 + \frac{C(1+b)}{k} - \frac{\omega^2 m}{k} \right) g(n).
\end{equation}
Considering this quantity on the right-hand side as a pseudo-energy $\mathcal{E}= 2 + (1+b)/k - \omega^2 m/k$, we see that this equation is equivalent to Harper's equation \eqref{eq:harper}.

%\todo[inline]{what are $E$ and $I$ in the formula $\tilde K_n=3E_n I_n/l^3$?} 

The spectral radius of \eqref{eq:diff2} is known to be $2+C/k$, meaning that 
$$-2-\frac{C}{k}\leq 2 + \frac{C(1+b)}{k} - \frac{\omega^2 m}{k}\leq 2+\frac{C}{k}.$$
As a result, we have that
\begin{equation}
    \frac{Cb}{m}\leq \omega^2 \leq \frac{4k+C(2+b)}{m}.
\end{equation}
As a result, there is always a low-frequency band gap up to at least $\omega=\sqrt{Cb/m}$ (in practice, this low-frequency band gap will often be much larger than this estimate, as we will see below).

This system is designed to be as simple as possible but have a spectrum that is equivalent to Hofstadter's butterfly. Masses coupled by linear springs is the natural candidate for toy models of waves in periodic (or modulated periodic) media (see \emph{e.g.} \cite{brillouin1953wave} and any subsequent undergraduate textbook). In particular, mass-spring systems provide immediate analogues of tight-binding Hamiltonians. However, such a system is not sufficient to obtain Harper's equation, as demonstrated in Appendix~\ref{app:matricitransferimento}. In order to be able to obtain the correct on-site modulation in the difference equation, we needed to introduce local dispersion to the system. We chose to achieve this through the addition of the Euler-Bernoulli beams, which introduce local resonances.

\section{Bloch spectra for rational modulation parameters}

If the parameter $\phi$ in the modulation \eqref{eq:modulation} is rational, then the oscillations of $\cos(2\pi\phi n)$ are commensurate with the periodic lattice. As a result, a repeating unit cell can be identified and a Bloch ansatz applied to find the spectrum. In particular, if $\phi=q/L$ where $q$ and $L$ are coprime integers, then the function  $n\mapsto\cos(2\pi\phi n)$ will be periodic with period $L$. We will denote this period as $L(\phi)$ to emphasise that it depends on $\phi$.

To compute the Bloch spectrum, it is convenient to rewrite the difference equation \eqref{eq:diff2} as a matrix equation
\begin{equation} \label{eq:Mn}
    \begin{pmatrix} g(n+1) \\ g(n)\end{pmatrix}
    =M_n(\omega)
    \begin{pmatrix} g(n) \\ g(n-1)\end{pmatrix}
    \quad\text{where}\quad 
    M_n(\omega)=\begin{pmatrix}
        2 + \frac{C(1+b)}{k} - \frac{\omega^2 m}{k} -\frac{C}{k}\cos(2\pi\phi n) & -1 \\ 1 & 0
    \end{pmatrix}.
\end{equation}
Hence, if we have a configuration that is periodic with period $L(\phi)$, then we can write 
\begin{equation} \label{eq:TM}
    \begin{pmatrix} g(n+L(\phi)) \\ g(n+L(\phi)-1)\end{pmatrix}
    =\mathcal{M}_{L(\phi)}(\omega)
    \begin{pmatrix} g(n) \\ g(n-1)\end{pmatrix}
    \quad\text{where}\quad 
    \mathcal{M}_{L(\phi)}(\omega)=\prod_{i=0}^{L-1} M_{n+i}(\omega).
\end{equation}
The Bloch modes of the system are those which satisfy $g(n+L(\phi))=\exp(\ii L(\phi) \kappa)g(n)$ for all $n$, where the Bloch momentum $\kappa$ is a real-valued parameter. Comparing this ansatz with \eqref{eq:TM} we see that $\det(\mathcal{M}_{L(\phi)}(\omega)-\exp(\ii L(\phi) \kappa)I)=0$. Using the fact that $\det(\mathcal{M}_{L(\phi)})=1$, this is equivalent to the dispersion relation
\begin{equation} \label{eq:dispersion}
    \cos(L\kappa)=\frac{1}{2}\tr(\mathcal{M}_{L(\phi)}(\omega)).
\end{equation}

\begin{figure}
    \centering
    \includegraphics[width=\linewidth]{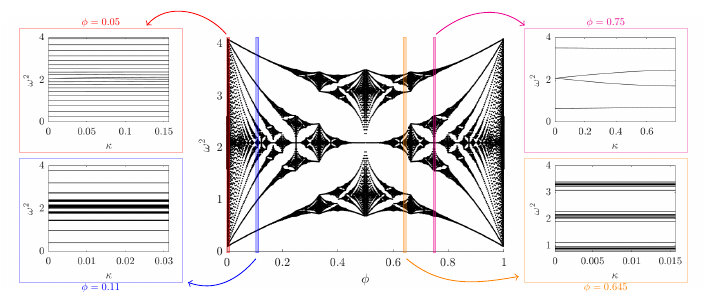}
    \caption{Hofstadter butterfly in the Bloch spectra of a periodic system with a rational modulation parameter. Here, we take unit masses with $m=1$, spring constants $k=0.5$ and modulation according to \eqref{eq:modulation} with parameter values $b=0.1$ and $C=1$. In the central plot, Bloch spectra are computed for the  rational values $\phi=0,1/400,2/400,\dots,399/400,1$. The four sub-plots show the dispersion curves for four specific values of $\phi$.}
    \label{fig:butterfly_bloch}
\end{figure}

We can use this formula \eqref{eq:dispersion} to generate a Hofstadter butterfly image, shown in Figure~\ref{fig:butterfly_bloch}. We consider the the rational values of $\phi$. For each value of $\phi$, we can compute the matrix $\mathcal{M}_{L(\phi)}(\omega)$ using the product in \eqref{eq:TM} and then use the dispersion relation \eqref{eq:dispersion} to compute dispersion curves. Some exemplar dispersion curves are shown for $\phi=0.05$, $\phi=0.11$, $\phi=0.75$ and $\phi=0.645$ in the sub-plots in Figure~\ref{fig:butterfly_bloch}. The corresponding lengths of the unit cells are, respectively, $L(0.05)=20$, $L(0.11)=100$, $L(0.75)=4$ and $L(0.645)=200$. From these examples in Figure~\ref{fig:butterfly_bloch} it is clear that the Bloch spectrum becomes increasingly intricate for larger $L(\phi)$ as the number of spectral bands increases accordingly. Note also that the width of the irreducible Brillouin zone $[0,\pi/L(\phi)]$ is scaled by $L(\phi)$, which is why the bands appear very flat in cases where $L(\phi)$ is large.

The central plot in Figure~\ref{fig:butterfly_bloch} is the result of combinging the dispersion curves for each value of $\phi=0,1/400,2/400,\dots,399/400,1$. For each value of $\phi$, we plot all the values of $\omega^2$ which lie somewhere on a dispersion curve (the associated value of the Bloch parameter $\kappa$ is not accounted for in this plot).

In some cases, the right-hand side of the dispersion relation \eqref{eq:dispersion} can be simplified. For example, when $\phi=0$, the system is a singly periodic array of masses and springs mounted on beams and the equation \eqref{eq:diff2} becomes
\begin{equation}
    g(n+1)+g(n-1)+\frac{C}{k}g(n) = \left( 2 + \frac{C(1+b)}{k} - \frac{\omega^2 m}{k} \right) g(n).
\end{equation}
We know the solutions must have the form $g(n)=\exp(\ii \kappa n)$, which leads to the dispersion relation
\begin{equation} \label{eq:single_dispersion}
    \omega^2=\frac{k}{m}\left( 2+\frac{C b}{k}-2\frac{C}{k}\cos(\kappa) \right).
\end{equation}
In general, however, such concise formulas relating $\omega$ and $\kappa$ explicitly are not tractable so the Hofstadter butterfly shown in Figure~\ref{fig:butterfly_bloch} must be obtained numerically from \eqref{eq:dispersion}. Studying \eqref{eq:single_dispersion}, we see that the phononic system has a high-frequency band gap above the critical frequency $\omega^2=(2k+C(2+b))/m$. Any such singly periodic system will have a similar high-frequency band gap, so we can directly deduce that any more complicated system with a larger unit cell will inherit a similar high-frequency band gap \cite{dunckley2024hierarchical, hori1964structureI}.

\section{Discrete spectra for finite structures}

Let us now consider the case where we have a finite-sized piece of the mechanical system consisting of $N+2$ masses. Suppose that we fix the position of the end masses, such that their displacement is always zero. Then, we can write that $g(0)=g(N+1)=0$ and we are left with a system of $N$ equations for the displacements of the $N$ free masses, as specified in \eqref{eq:diff2}. Thus, the eigenvalues $\omega^2$ can be calculated easily as the eigenvalues of a tri-diagonal matrix. Further, in this setting it is straightforward to ask questions about the nature of the associated eigenvectors and whether they correspond to energy localisation or propagation.

The localisation of the finite-sized system's eigenvectors can be quantified using the inverse participation ratio (IPR) \cite{biddle2012IPR}, defined as  
\begin{equation} \label{eq:IPR}
    \mathrm{IPR} = \frac{\sum_{n=1}^N |g(n)|^4}{\left( \sum_{n=1}^N |g(n)|^2 \right)^2}.
\end{equation}
For any eigenvector of the considered finite system, corresponding to an eigenvalue $\omega^2$, the quantity \eqref{eq:IPR} indicates the inverse of the number of the occupied sites in the structure, so when the oscillations are equally distributed and all the masses are vibrating, we have $\mathrm{IPR} \approx 1/N$. This corresponds to energy propagating through the system. In the opposite situation of extreme localisation, when only one site is vibrating with the associate frequency, we have $\mathrm{IPR} \approx 1$. Consequently, for structures composed of a large number of sites $N$, the $\mathrm{IPR}$ take values in the interval $[0,1]$. A small value indicates that the vector $(g(n))$ is delocalised, meaning the eigenmode propagates wave energy through the system, whereas a value close to 1 corresponds to wave energy being localised.

\begin{figure}
    \centering
    \begin{subfigure}{\linewidth}
    \centering
        \includegraphics[width=0.8\linewidth]{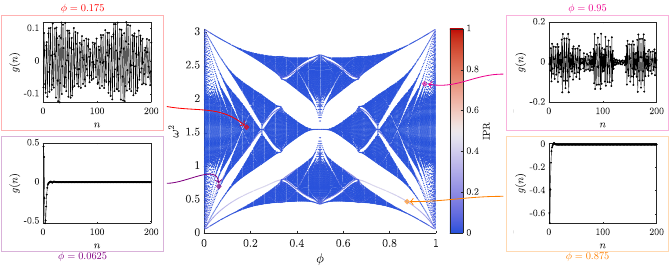}
        \caption{$C/k=1$}
    \end{subfigure}
    \begin{subfigure}{\linewidth}
    \centering
        \includegraphics[width=0.8\linewidth]{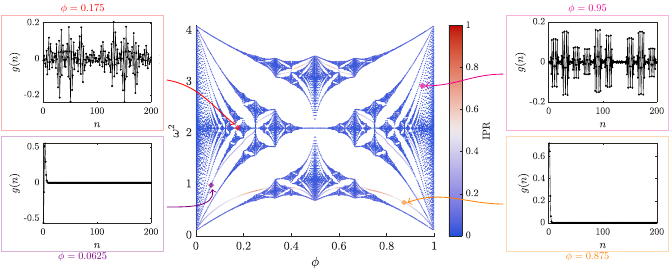}
        \caption{$C/k=2$}
    \end{subfigure}
    \begin{subfigure}{\linewidth}
    \centering
        \includegraphics[width=0.8\linewidth]{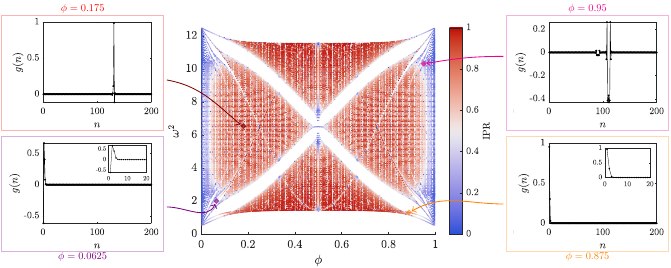}
        \caption{$C/k=10$}
    \end{subfigure}

    \hspace{0.05cm}
    
    \begin{subfigure}{\linewidth}
        \centering
        \includegraphics[width=0.45\linewidth]{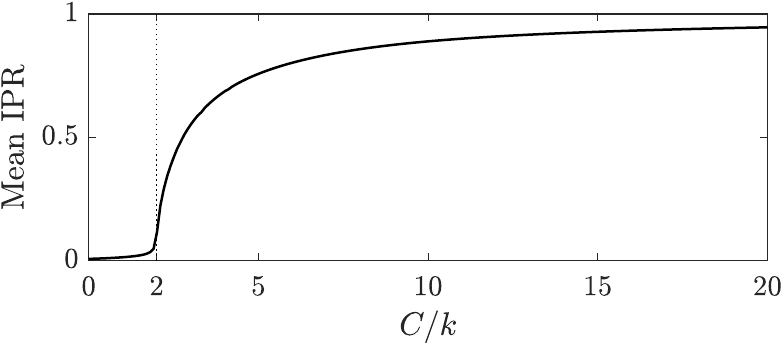}
        \caption{Metal-insulator transition ($\phi = (1+\sqrt{5})/2$)}
    \end{subfigure}
    \caption{Hoftstadter butterfly in the discrete set of eigenvalues of a finite-sized structure with $N=200$ free massed and fixed end conditions. Here, we take unit masses with $m=1$, spring constants $k=0.5$ and modulation according to \eqref{eq:modulation} with parameter values $b=0.1$ and either (a) $C=0.5$, (b) $C=1$ or (c) $C=5$. In each case, the central plots show the discrete spectra computed for the rational values $\phi=0,1/400,2/400,\dots,399/400,1$. The color of each point indicates the inverse participation ratio (IPR) of the associated eigenvector. Additionally, four exemplar eigenvectors have been picked out, show in the inset plots. Finally, (d) shows how the mean IPR (averaged over all 200 eigenvectors) varies as a function of $C/k$, for fixed $\phi=(1+\sqrt{5})/2$.}
    \label{fig:butterfly_finitefixed}
\end{figure}

The strong dependence of the value of the $\mathrm{IPR}$ on the ratio $C/k$ and the implications on the layout of the Hofstadter butterfly spectra are illustrated by the numerical results reported in Figure~\ref{fig:butterfly_finitefixed}. The butterfly spectra shown in this figure are associated to finite-sized structures composed of $N=200$ free masses with $m=1$, $k=0.5$, $b=0.1$ and different values of $C$. We considered $C=0.5$ corresponding to $C/k=1$ (Figure~\ref{fig:butterfly_finitefixed}$(a)$), $C=1$ yielding $C/k=2$ (Figure~\ref{fig:butterfly_finitefixed}$(b)$) and finally $C=5$ associated with $C/k=10$ (Figure~\ref{fig:butterfly_finitefixed}$(c)$). We can clearly see that for $C/k=1$ the spectrum possesses well defined bands and almost all eigenvalues $\omega^2$ correspond to eigenvectors with $\mathrm{IPR}$ values close to zero, as indicated by the color map. Only the few edge states that cross the larger band gaps are localised, such as those corresponding to $\phi = 0.0625$ and $\phi=0.875$ reported in the plots at the sides of the spectrum. 

For $C/k=2$, we can note that the larger and better defined band gaps detected assuming $C=/k=1$ become even wider and new gaps with similar shape appear between them. This is clearly a self-similar (multi-fractal) behaviour governed by the variation of the modulation amplitude $C/k$, and it is in agreement to what observed in several comparable systems \cite{rosa2021exploring, comi2024some, richoux2002acoustic, ni2019observation, martinez2018quasiperiodic}. Also for this case, the localised states are associated with the eigenvalues crossing the larger gaps, as shown by the eigenvectors plotted in Figure~\ref{fig:butterfly_finitefixed}$(b)$. As we will see further in Figure~\ref{fig:butterfly_finitefixed}$(d)$, $C/k=2$ is a critical value for the amplitude of the modulation, associated with a change in the dynamical properties of the system.

Figure~\ref{fig:butterfly_finitefixed}$(c)$ shows that as $C/k$ increases further and reaches for example $C/k=10$, the larger gaps lose their characteristic shapes, and we have a transition to an almost unbanded spectrum. Indeed, for this case the states appears in very narrow sets such that it is difficult to distinguish band gaps and propagation curves for most of the domain except the four main large gaps. As indicated by the colour map, most of the eigenvalues correspond to eigenvalues with $\mathrm{IPR}$ close to $1$, and then to localised states as those associated with $\phi=0.0625, 0.175, 0.95, 0.875$ reported on the sides of the spectrum.

The behaviour of the spectrum illustrated in Figures~\ref{fig:butterfly_finitefixed}$(a)$, $(b)$ and $(c)$ is a demonstration of the phase transition that we can observe in this system, governed by varying the amplitude of the modulation $C/k$. In order to study this phenomenon, in Figure~\ref{fig:butterfly_finitefixed}$(d)$ we present the inverse participation ratio averaged over all the eigenvectors (mean IPR) as a function of the ratio $C/k$. In this case, we have $200$ eigenvalues corresponding to the number of oscillating masses composing the considered finite-sized structure. For $C/k<2$, the mean IPR varies slowly and has values much smaller than 1, of the order of $10^{-1}$, while for $C/k>2$ it starts to grow much faster until it asymptotically reaches the value 1. This means that for $C/k<2$ most of the eigenvectors have an IPR much smaller than $1$, corresponding to non-localised modes associated with the vibration of many of the masses composing the system (see for example the cases with $\phi=0.175$ and $\phi=0.95$ in Figures~\ref{fig:butterfly_finitefixed}$(a)$ and $(b)$). Conversely, for $C/k>2$ the number of eigenvectors with higher IPR increases until almost all of them have $\text{IPR} \approx 1$, and then they correspond to extremely localised states characterized by the vibration of a single or a few nearby masses (see for example the four cases shown in Figures~\ref{fig:butterfly_finitefixed}$(c)$). We can deduce from this analysis that $C/k=2$ represents a critical value of the stiffness modulation amplitude, and that for $C/k>2$ the dynamical state of the system changes from mainly extended to localised, similarly to the metal-insulator phase transition observed in several one-dimensional structures investigated in condensed matter physics \cite{biddle2012IPR, aulbach2004phase}.

The observed metal-insulator transition is in agreement to what predicted by the rigorous spectral analysis of systems described by similar differential operators \cite{aubry1980analyticity, jitomirskaya1999metal}. It could have important applications for designing structures with tunable mechanical properties which can switch from transmitting vibrations in selected intervals of frequencies to localising them and storing their energy.

% \todo[inline]{Do different boundary conditions (how do we model free BCs at the ends?. Look at the topological properties of the edge modes.}

\section{Transmission spectra for finite structures}

We now compute the transmission coefficient for a finite-sized portion of the introduced mechanical system analogous to the one considered in the previous Section by adopting the approach set out in \cite{markos2008wave}. We show that by varying the vibration frequency as well as the modulation parameter $\phi$, the layout of the transmission coefficient reproduces the butterfly spectrum reported in Figure~\ref{fig:butterfly_finitefixed}.

Let us decompose the field at mass $n$ into waves propagation from left to right and from right to left, giving the expression
% recall the matrix equation \eqref{eq:TM}, which relates the displacement at masses $n = 0$ and $n = 1$ to the displacement at masses %$n = L$ and $n = L + 1$, through the matrix $\mathcal{M}_L$.
\begin{equation}\label{eq:wavedef}
    g(n) = A_n \exp(\ii \kappa n) + B_n \exp(-\ii \kappa n)= g^+(n)+g^-(n),
\end{equation}
for some wave number $\kappa$. The superscripts $+$ and $-$ indicate the components propagating in positive (from left to right) and negative (from right to left) direction, respectively. We can write that
\begin{equation} \label{eq:Qn}
\begin{pmatrix} g(n+1) \\ g(n) \end{pmatrix}
=
Q(\kappa)\begin{pmatrix} A_n\exp(\ii\kappa(n+1)) \\ B_n\exp(-\ii\kappa(n+1)) \end{pmatrix} \quad
\mathrm{and} \quad
\begin{pmatrix} g(n) \\ g(n-1) \end{pmatrix}
=
Q(\kappa)\begin{pmatrix} A_n\exp(\ii\kappa n) \\ B_n\exp(-\ii\kappa n) \end{pmatrix},
\end{equation}
where $Q(\kappa)$ is the matrix
\begin{equation}
    Q(\kappa)=\begin{pmatrix} 1 & 1 \\ \exp(-\ii\kappa) & \exp(\ii\kappa) \end{pmatrix}.
\end{equation}
We now adapt the matrix equation \eqref{eq:TM} to study the case of a finite-sized non periodic structure composed of $N$ masses. Specializing the expressions \eqref{eq:Qn}$_{(1)}$ and \eqref{eq:Qn}$_{(2)}$ respectively for $n=N$ and $n=1$ and substituting them into \eqref{eq:TM}, we obtain
\begin{equation} \label{eq:TransM}
    \begin{pmatrix} A_N \exp(\ii\kappa (N+1)) \\ B_N\exp(-\ii\kappa (N+1)) \end{pmatrix}
    =Q(\kappa)^{-1}\mathcal{M}_N(\omega)Q(\kappa)
    \begin{pmatrix} A_1 \exp(\ii\kappa) \\ B_1\exp(-\ii\kappa) \end{pmatrix}, \quad 
    \mathrm{where}
   \quad 
    \mathcal{M}_{N}(\omega)=\prod_{n=1}^{N} M_{n}(\omega). 
\end{equation}

The matrix $T_N=Q(\kappa)^{-1}\mathcal{M}_N(\omega)Q(\kappa)$ in \eqref{eq:TransM} is the transfer matrix of the section of length $N$: it relates left- and right-propagating waves at the first mass with those at the $(N+1)$\textsuperscript{th} mass. Assuming that the first mass is located at the left-boundary of the finite sample while the $(N+1)$\textsuperscript{th} is at the right end and  using the definition \eqref{eq:wavedef}, we introduce the notation $g_l^+=g^+(1)$, $g_l^-=g^-(1)$, $g_r^+=g^+(N+1)$, $g_r^-=g^-(N+1)$. Equation \eqref{eq:TransM} can be written in the form 
\begin{equation} \label{eq:TransMg}
    \begin{pmatrix} g^+_r \\ g^-_r \end{pmatrix}
    =T_N
    \begin{pmatrix} g^+_l \\ g^-_l \end{pmatrix}.
\end{equation}
For calculating a transmission coefficient, it is useful to rephrase the \eqref{eq:TransMg} in terms of a scattering matrix, which relates outgoing and incoming waves. The appropriate relation is
\begin{equation} \label{eq:scatt}
    \begin{pmatrix}  g^-_l \\ g^+_r
    \end{pmatrix}
    =S_N
    \begin{pmatrix} g^+_l \\ g^-_r \end{pmatrix},
\end{equation}
where due to the unimodularity property of $M_N$ ($\det M_N=1$), the scattering matrix $S_N$ is given by
\begin{equation}
    S_N=\begin{pmatrix}
        -\frac{(T_N)_{21}}{(T_N)_{22}} & \frac{1}{(T_N)_{22}} \\
        \frac{1}{(T_N)_{22}} & \frac{(T_N)_{12}}{(T_N)_{22}}.
    \end{pmatrix}
\end{equation}
Then, if we assume that there is no wave incident from the right (meaning that $g^-_l=0$), we can see that the transmitted field is given by
\begin{equation} \label{eq:t}
    t=\frac{g^+_r}{g^+_l}=(S_N)_{21}=\frac{1}{(T_N)_{22}},
\end{equation}
and the reflected field is
\begin{equation} \label{eq:r}
    r=\frac{g^-_l}{g^+_l}=(S_N)_{11}=-\frac{(T_N)_{21}}{(T_N)_{22}}.
\end{equation}
Finally, the reflection and transmission coefficients are given by $|r|^2$ and $|t|^2$, respectively. 

We now use the formulas \eqref{eq:t} and \eqref{eq:r} to compute transmission spectra for finite-sized quasiperiodically modulated arrays including $N$ masses. On either side of the section of length $N$, we assume that it is attached to infinite, singly periodic arrays (\emph{i.e.} the $\phi=0$ structure). Within the periodic structure, given an operating frequency $\omega$, we can use the dispersion relation \eqref{eq:single_dispersion} to calculate the wave number $\kappa$ (which is used in $Q(\kappa)$ when calculating $T_N$).

\begin{figure}
    \centering
    \includegraphics[width=0.9\linewidth]{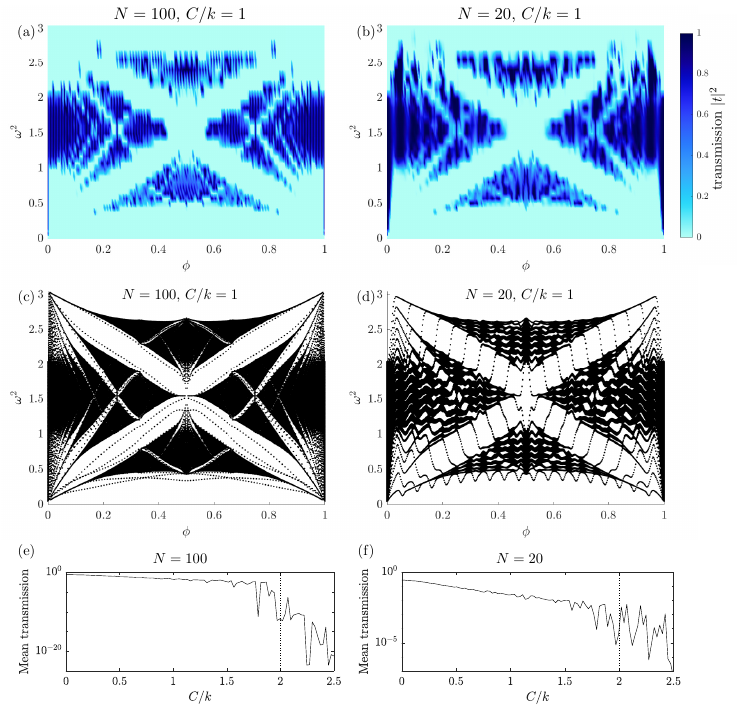}
    \caption{Hofstadter butterfly in the transmission coefficient of a finite-sized system. Here, we take unit masses with $m=1$, spring constants $k=0.5$ and modulation according to \eqref{eq:modulation} with parameter values $b=0.1$ and $C=0.5$. As a result, we have $C/k=1$ so the eigenvectors are generally delocalised. For each value of $\phi$ and $\omega$, we plot the transmission $|t|^2$, as in \eqref{eq:t}. (a) shows a system with $N=100$ masses whereas (b) shows a smaller system with $N=20$ masses. For comparison, (c) and (d) show the Bloch spectra of the same finite-sized systems with periodic boundary conditions. Finally, (e) and (f) show the mean transmission (averaged over the full frequency range $0\leq\omega^2\leq3$) as a function of $C/k$ for  fixed $\phi=(1+\sqrt{5})/2$.}
    \label{fig:transmission}
\end{figure}

The resulting transmission spectra are shown in Figure~\ref{fig:transmission}. By varying the phase parameter $\phi$ we see the Hofstadter butterfly emerge. We show the transmission coefficient for sections of length $N=100$ and $N=20$ in Figures~\ref{fig:transmission}(a) and \ref{fig:transmission}(b), respectively. We can see that the resolution of the butterfly increases in larger systems. It is important to note that we have chosen parameter values such that $C/k=1$ which is below the critical threshold metal-insulator transition ($C/k=2$), such that the eigenmodes of the system propagate energy. If we assume $C/k>2$, then the modes are exponentially localised meaning that the transmission is small at all frequencies and we do not see the butterfly emerge. This behaviour is highlighted by Figures~\ref{fig:transmission}(e) and \ref{fig:transmission}(f), which show how the mean transmission coefficient (averaged over all the frequencies in the range $0\leq\omega^2\leq3$) decreases quickly as $C/k$ is increased. We can observe that, in agreement to what detected in previous Section, the decrease of the mean transmission coefficient becomes particularly evident for $C/k>2$, until at $C/k=2.5$ it becomes of the order $10^{-20}$ for $N=100$ and approximately $10^{-7}$ for $N=20$. This confirms that for $C/k>2$ almost all the frequencies correspond to localised modes and the considered finite-sized structure behaves like an insulator for the mechanical vibrations. For comparison, in Figures~\ref{fig:transmission}(c) and \ref{fig:transmission}(d) we show the butterflies obtained by computing the Bloch spectra with the same sections taken as the unit cell. The change in resolution is clear: from the $N=100$ system we can clearly see the fractal collection of gaps, but this is somewhat blurred in the $N=20$ system. In both cases, we have some pollution with extra eigenvalues appearing in the band gaps (compared to what we expect for Bloch spectra, based on Figure~\ref{fig:butterfly_bloch}). This is due to the fact that for many of the values of $\phi$ used in these calculations, although the full system is periodic its unit cell is much larger than 20 or 100 masses, so we are artificially cutting the unit cell, which has the effect of slightly distorting the spectrum away from the classical Hofstadter butterfly.

\section{Concluding remarks}

In this work, we have devised a one-dimensional phononic system that produces the Hofstadter butterfly exactly. This system is composed of simple discrete elements where the crucial modulation can be achieved by varying quasiperiodically the thickness of the beams on which the masses are mounted. We surveyed the various different ways of achieving the butterfly (Bloch spectra, eigenvalues of finite-sized samples and transmission spectra) and highlighted how they achieve differing resolutions and varying tendencies to have spectra polluted by edge modes. We found that the frequency spectrum can be strongly modified according to the variation of amplitude of the modulation and we determined a critical value for this parameter, associated with a change in the dynamical state of the system from mainly extended to localised, in analogy with the metal-insulator phase transitions investigated in condensed matter physics. This phase transition is highlighted also in the analysis of the transmission coefficient, representing for this simple vibro-mechanical structure the easiest way to generate experimental data. Our numerical results show how even considering  a relatively short array ($20$ masses) the transmission spectra layout gives a reproduction of the butterfly. This provides the possibility to design relatively easy-to-implement systems which can be alternatively transparent (in the sense that the eigenmodes propagate energy) and opaque (in the sense that most of the modes are localised) according to different amplitude of the modulation of the beams thickness. These results could open new perspectives towards the realisation of phononic structures whose behaviour can range from that of a sort of conductor to an insulator-like device for what concerns the propagation of mechanical vibrations.

The Harper-type modulation considered here is just one of the quasiperiodic patterns studied in the literature. The other widely studied one-dimensional class of patterns are the so-called (generalised) Fibonacci tilings, which are known to have fractal spectra \cite{kohmoto1983localization, kola1990one, davies2024super, jagannathan2021fibonacci} and haven been shown to have some topological equivalence with the Hofstadter butterfly \cite{naumis2008electronic, kraus2012topological}. In two or more dimensions, there is an exciting variety of quasiperiodic patterns, which are largely unexplored for realising novel metamaterial phenomena \cite{beli2021mechanics, zhang2023chiral}.

\section*{Acknowledgments}

The authors would like to thank Richard Craster and Gregory Chaplain for helpful discussions on computing transmission coefficients for discrete systems. B.D. was partly supported by the EPSRC under grant number EP/X027422/1. L.M. was supported by the Italian Government through PNRR project AISAC-B29J2300112005 (‘finanziato dal Piano Nazionale di Ricrescita e Resilienza’).

\section*{Data availability}

The Matlab codes used to produce the plots presented in this article are available for download at \url{https://doi.org/10.5281/zenodo.15010829}.

\appendix
\section{General Hofstadter butterfly conditions} \label{app:matricitransferimento}

In this appendix, we derive general conditions for a one-dimensional system to have a spectrum that is equivalent to the Hofstadter butterfly. This will show that simpler versions of the system considered in this work, for example with the local resonances (due to the beams) removed, do not have sufficient complexity to be able to recover Harper's equation. Similarly, the modulation needs to be on these local resonances (as opposed to on the coupling strengths or, equivalently, on the spacing between the resonant elements).

Suppose we have a one-dimensional system that is fully characterised by a two-element state vector 
\begin{equation}
    \mathbf{u}_n = \begin{pmatrix} A_n \\ B_n \end{pmatrix}.
\end{equation}
Further, suppose that the propagation of waves through this system is described by a $2\times2$ unimodular (\emph{i.e.} having determinant equal to 1) transfer matrix $T_n$, which is allowed to depend on the position index $n$, in the sense that
\begin{equation} \label{eq:transfer}
    \mathbf{u}_{n+1} = T_n\mathbf{u}_n.
\end{equation}
We can relate this system to Harper's equation \eqref{eq:harper}. It holds that
\begin{equation} \label{eq:1}
    A_{n+1} = (T_n)_{11} A_{n} + (T_n)_{12} B_{n}.
\end{equation}
Further, by inverting \eqref{eq:transfer} we have that
\begin{equation} \label{eq:2}
    A_{n-1}=(T_{n-1})_{22} A_{n} - (T_{n-1})_{12} B_{n}.
\end{equation}
Adding \eqref{eq:1} and \eqref{eq:2}, we find that 
\begin{equation} \label{eq:maybeHarper}
    A_{n+1}+A_{n-1}=\left[(T_n)_{11}+(T_{n-1})_{22}\right]A_n+\left[(T_n)_{12}-(T_{n-1})_{12}\right]B_n.
\end{equation}
Thus, we can see that if a system has a unimodular transfer matrix $T_n$ that satisfies
\begin{equation} \label{eq:condition1}
    (T_n)_{11}+(T_{n-1})_{22}=E-2\cos(2\pi\phi n),
\end{equation}
for all $n\in\mathbb{Z}$, where $E$ is a real-valued pseudo-energy (that does not depend on $n$), and
\begin{equation} \label{eq:condition2}
    (T_n)_{12}-(T_{n-1})_{12}=0,
\end{equation}
for all $n\in\mathbb{Z}$, then the equation \eqref{eq:maybeHarper} reduces to Herper's equation \eqref{eq:harper}, meaning that the system has a spectrum that is equivalent to the Hofstadter butterfly. An interpretation of these conditions is that \eqref{eq:condition1} means the ``on-site'' modulation follows the Harper-type law and \eqref{eq:condition2} means the coupling between elements is kept constant. 

Consider a simplified version of the system considered in this work, which is an array of masses coupled with springs. Suppose that the $n$\textsuperscript{th} mass has mass given by $m_n$ and the $n$\textsuperscript{th} spring has spring constant $k_n$. We consider a time-harmonic wave propagating along this array and let $u_ne^{i\omega t}$ be the displacement of the $n$\textsuperscript{th} mass and $f_ne^{i\omega t}$ be the harmonic force acting on that mass. Then, the transfer matrix for this system is given by \cite{Lazaro2022Sturm}
\begin{equation} \label{eq:massspring}
\begin{pmatrix} u_n \\ f_n \end{pmatrix}=
\begin{pmatrix}
1 & -\cfrac{1}{k_n} \\
m_n\omega^2 & 1-\cfrac{m_n\omega^2}{k_n} \\
\end{pmatrix}
\begin{pmatrix} u_{n-1} \\ f_{n-1} \end{pmatrix}.
\end{equation}
Clearly, for the coupling condition \eqref{eq:condition2} to hold, we must have all the springs be identical, so we need to take $k_n=k$ for all $n$. Then, to meet the on-site condition \eqref{eq:condition1}, it must be the case that
\begin{equation}
    2-\cfrac{m_{n-1}\omega^2}{k}=E-2\cos(2\pi\phi n).
\end{equation}
This can only be possible if $m_{n-1}$ or $k$ depend on $\omega$, which is not the case in a simple mass-spring system.

The most straightforward way to achieve frequency-dependent local properties in a mechanical system is through the introduction of local resonances, which was the motivation introducing the Euler-Bernoulli beams in the system considered in this work. In this case, we can (somewhat trivially) write the difference equation \eqref{eq:diff2} as a 2-by-2 system
\begin{equation}
    \begin{pmatrix} g(n+1) \\ g(n)\end{pmatrix}
    =\begin{pmatrix}
        4 + \epsilon - \frac{\omega^2 m}{K} -2\cos(2\pi\phi n) & -1 \\ 1 & 0
    \end{pmatrix}
    \begin{pmatrix} g(n) \\ g(n-1)\end{pmatrix}.
\end{equation}
In which case, it is immediately clear that the conditions \eqref{eq:condition1} and \eqref{eq:condition2} are satisfied, subject to the appropriate choice of the pseudo-energy $E$.

\bibliographystyle{ieeetr}
\bibliography{references}{}

\end{document}